\documentclass[aps, prd, amsmath, floats, floatfix, twocolumn, nofootinbib, superscriptaddress]{revtex4-1}
\usepackage[pdftex]{graphicx}
\usepackage{color}
\usepackage{latexsym}
\usepackage[colorlinks]{hyperref}

\newcommand{\be}{\begin{eqnarray}}
\newcommand{\ee}{\end{eqnarray}}

\begin{document}

\title{Testing the Weak Equivalence Principle near black holes}

\author{Rittick~Roy}
\affiliation{Center for Field Theory and Particle Physics and Department of Physics, Fudan University, 200438 Shanghai, China}

\author{Askar~B.~Abdikamalov}
\affiliation{Center for Field Theory and Particle Physics and Department of Physics, Fudan University, 200438 Shanghai, China}
\affiliation{Ulugh Beg Astronomical Institute, Tashkent 100052, Uzbekistan}

\author{Dimitry~Ayzenberg}
\affiliation{Theoretical Astrophysics, Eberhard-Karls Universit\"at T\"ubingen, D-72076 T\"ubingen, Germany}

\author{Cosimo~Bambi}
\email[Corresponding author: ]{bambi@fudan.edu.cn}
\affiliation{Center for Field Theory and Particle Physics and Department of Physics, Fudan University, 200438 Shanghai, China}

\author{Shafqat~Riaz}
\affiliation{Center for Field Theory and Particle Physics and Department of Physics, Fudan University, 200438 Shanghai, China}

\author{Ashutosh~Tripathi}
\affiliation{Center for Field Theory and Particle Physics and Department of Physics, Fudan University, 200438 Shanghai, China}

\begin{abstract}
Today we have quite stringent constraints on possible violations of the Weak Equivalence Principle from the comparison of the acceleration of test-bodies of different composition in Earth's gravitational field. In the present paper, we propose a test of the Weak Equivalence Principle in the strong gravitational field of black holes. We construct a relativistic reflection model in which either the massive particles of the gas of the accretion disk or the photons emitted by the disk may not follow the geodesics of the spacetime. We employ our model to analyze the reflection features of a \textsl{NuSTAR} spectrum of the black hole binary EXO~1846--031 and we constrain two parameters that quantify a possible violation of the Weak Equivalence Principle by massive particles and X-ray photons, respectively.
\end{abstract}

\maketitle


\section{Introduction}

The theory of general relativity was proposed by Einstein at the end of 1915~\cite{Einstein:1916vd}. After more than a century and without any modification, it is still a pillar of modern physics, representing our standard framework for the description of gravitational fields and the spacetime structure. The theory has been extensively tested in weak gravitational fields since the 1960s with solar system experiments and since the 1970s with radio observations of binary pulsars~\cite{Will:2014kxa}. However, thanks to a new generation of observational facilities, the past five years have seen a tremendous progress in our capabilities of testing general relativity in the strong field regime and we can now verify the predictions of Einstein's gravity around black holes with gravitational waves~\cite{TheLIGOScientific:2016src,Yunes:2016jcc,LIGOScientific:2019fpa}, mm VLBI data~\cite{Davoudiasl:2019nlo,Bambi:2019tjh,Psaltis:2020lvx}, and X-ray observations~\cite{Cao:2017kdq,Tripathi:2018lhx,Tripathi:2020dni}.

Gravitational wave tests and electromagnetic tests can probe different sectors of the theory and are thus complementary~\cite{Yunes:2013dva,Bambi:2015kza,Yagi:2016jml}. Gravitational waves can directly test the dynamical regime and are more suitable to study the gravitational sector itself. Electromagnetic tests can probe better the interactions between the gravity and the matter sectors. For example, electromagnetic tests cannot distinguish Kerr black holes in different metric theories of gravity, while gravitational wave tests may do it~\cite{Psaltis:2007cw,Barausse:2008xv}. On the other hand, a non-minimal coupling between the matter and gravity sectors may leave a signature in the electromagnetic spectrum without affecting the gravitational wave signal~\cite{Bambi:2013mha,Li:2019lsm,Li:2021mzq}.

One of the postulates of general relativity and of all the metric theories of gravity is the validity of the Weak Equivalence Principle (WEP), which asserts that the trajectory of freely falling test-bodies is independent of their internal structure and composition~\cite{Will:2014kxa}. The WEP requires that all freely falling test-bodies follow the geodesics of the spacetime. Tests of the WEP normally compare the acceleration of two test-bodies in Earth's gravitational field and constrain the E\"otv\"os parameter $\eta = 2|a_1 - a_2|/|a_1 + a_2|$, where $a_1$ and $a_2$ are the accelerations of, respectively, body~1 and body~2. The current most stringent constraint on the E\"otv\"os parameter comes for the \textsl{MICROSCOPE} mission, which compares the accelerations of two masses of titanium and platinum alloys, respectively, inferring the measurement $\eta < 10^{-14}$~\cite{Touboul:2017grn}.

Violations of the WEP are common in extensions of general relativity or in theories aiming at unifying gravity with the other forces~\cite{Will:2014kxa}. For example, if we have a theory in which gravity is described by the spacetime metric $g_{\mu\nu}$ and the scalar field $\phi$, we can write the action of the gravity and matter sectors as
\be
\hspace{-0.5cm}
S = S_{\rm gravity} [ g_{\mu\nu} , \phi ] + S_{\rm matter} [ \psi_{m1} , \psi_{m2} , ... , g_{\mu\nu} , \phi ] \, ,
\ee
where \{$\psi_{mi}$\} are different matter fields. In the case of a metric theory of gravity, all matter fields respond to some metric
\be\label{eq-conf}
\tilde{g}_{\mu\nu} = A^2(\phi) g_{\mu\nu} \, ,
\ee
where $A(\phi)$ is a function of the scalar field $\phi$. If we perform the conformal transformation in Eq.~(\ref{eq-conf}), the total action can be written as
\be
S = S_{\rm gravity} [ g_{\mu\nu} , \phi ] + S_{\rm matter} [ \psi_{m1} , \psi_{m2} , ... , \tilde{g}_{\mu\nu} ] \, ,
\ee
and we find that all particles follow the geodesics of the metric $\tilde{g}_{\mu\nu}$, as it is required by the WEP. On the other hand, if the scalar field $\phi$ does not universally couple to matter, the action has the form
\be
S &=& S_{\rm gravity} [ g_{\mu\nu} , \phi ] + S_{\rm matter1} [ \psi_{m1} , g_{\mu\nu} , \phi ] \nonumber\\
&& + S_{\rm matter2} [ \psi_{m2} , g_{\mu\nu} , \phi ] + ... \, .
\ee
Even if the particles associated to the matter field $\psi_{m1}$ follow the geodesics of the metric $\tilde{g}_{\mu\nu}$, those associated to the other matter fields do not. The particles associated to the matter field $\psi_{m2}$ may follow the geodesics of the metric $\hat{g}_{\mu\nu} = B^2(\phi) g_{\mu\nu}$, those associated to the matter field $\psi_{m3}$ the geodesics of the metric $\bar{g}_{\mu\nu} = C^2(\phi) g_{\mu\nu}$, etc. Different particle species respond to different background metrics and the WEP does not hold.

In this paper, we propose a novel test of the WEP in the strong gravitational field of black holes using X-ray reflection spectroscopy. While this technique is not new for testing general relativity in the strong field regime, as of now it has only been employed to test the Kerr metric around black holes {\it assuming} the WEP (e.g.,~\cite{Cao:2017kdq,Tripathi:2018lhx,Tripathi:2020dni}). Here we assume the Kerr metric~\cite{Psaltis:2007cw} and we show that X-ray reflection spectroscopy can be used to test the WEP, in particular the possibility that either massive particles or X-ray photons do not follow the geodesics of the spacetime. Such a test is not possible with gravitational waves because the gravitational wave signal is independent of photon trajectories. Following an agnostic approach, we do not consider a specific theory but we simply assume that some particles follow the geodesics of the Kerr spacetime and other particles follow the geodesics of a deformed Kerr spacetime. We construct a reflection model for such a scenario and constrain the parameters associated to the non-geodesic motion by fitting a \textsl{NuSTAR} observation of the black hole binary EXO~1846--031.

The content of the paper is as follows. In Section~\ref{s-gr}, we briefly review the calculations of the reflection spectrum of a black hole in general relativity, where the WEP holds. In Section~\ref{s-ngm}, we present our theoretical framework to calculate reflection spectra of black holes in models violating the WEP. In Section~\ref{s-obs}, we apply our new method to test the WEP in the strong gravity region of the stellat-mass black hole in EXO~1846--031. We discuss our results in Section~\ref{s-dis}.


\section{Reflection spectrum in general relativity}\label{s-gr}

\begin{figure}[t]
\begin{center}
\includegraphics[width=8.5cm,trim={0cm 0cm 0cm 0cm},clip]{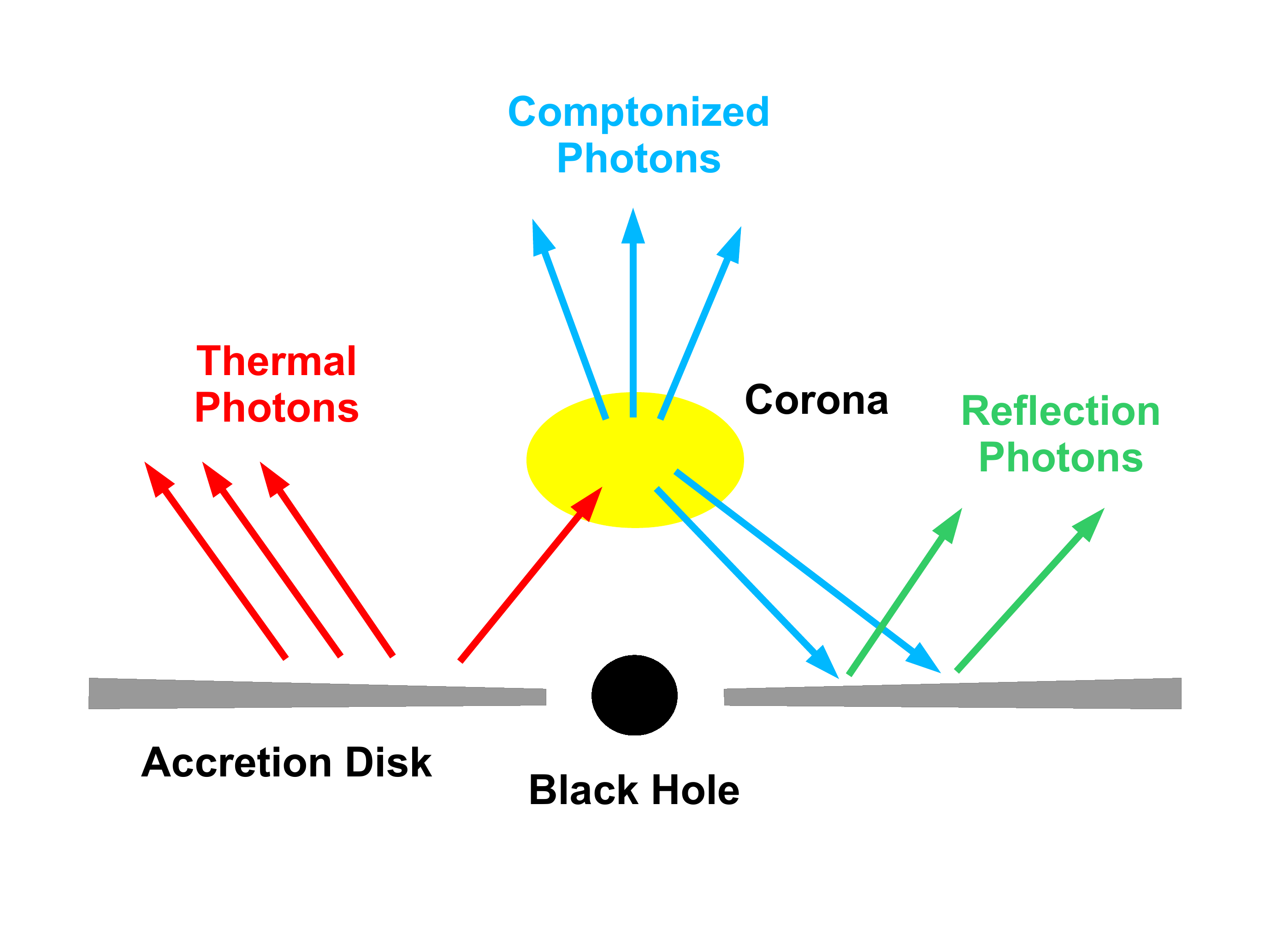}
\end{center}
\vspace{-1.2cm}
\caption{A black hole is surrounded by a geometrically thin and optically thick accretion disk. Thermal photons from the disk inverse Compton scatter off free electrons in the corona. A fraction of the Comptonized photons illuminate the disk, generating a reflection component. \label{f-corona}}
\end{figure}

Blurred reflection features are common in the X-ray spectra of accreting black holes~\cite{Nandra:2007rp,Walton:2012aw,Tripathi:2020yts} and are thought to be generated by illumination of the inner part of the accretion disk by a hot corona; see, e.g., \cite{Bambi:2020jpe} and references therein. The system is illustrated in Fig.~\ref{f-corona}. The accretion disk is geometrically thin and optically thick. Every point on the accretion disk has a blackbody-like spectrum and the whole disk has a multi-temperature blackbody-like spectrum. The emission of the inner part of the accretion disk is normally peaked in the soft X-ray band (0.1-1~keV) for stellar-mass black holes and in the UV band (1-100~eV) for supermassive black holes. The term ``corona'' denotes some hotter ($\sim$100~keV) plasma near the black hole. For example, the corona may be the atmosphere above the accretion disk, some gas between the inner part of the accretion disk and the black hole, or the base of a possible jet. Thermal photons from the accretion disk can inverse Compton scatter off free electrons in the corona. These Comptonized photons have a spectrum that can be approximated by a power-law with an exponential high-energy cutoff. A fraction of the Comptonized photons illuminate the disk, followed by Compton scattering, absorption, and fluorescent emission that produce the reflection component.

In the rest-frame of the gas, the reflection spectrum is characterized by a soft excess below 2~keV, narrow fluorescent emission lines below 10~keV, and a Compton hump peaked at 20-30~keV~\cite{Ross:2005dm,Garcia:2010iz}. For our purposes, the fluorescent emission lines are of particular interest as they are particularly influenced by the strong gravity of the black hole spacetime. Due to relativistic effects (light-bending, Doppler boosting, gravitational redshift) the lines become broadened and skewed after emission and while traveling through the black hole spacetime on the way to the observer~\cite{Fabian:1989ej,Laor:1991nc}. Ultimately, the reflection spectrum is dependent on the properties of the spacetime and can, in principle, be used to extract the black hole's physical parameters and study gravity in a strong field regime.

We consider a Kerr black hole with an infinitesimally thin accretion disk in the equatorial plane. The reflection spectrum of the disk is characterized by the specific intensity, $I_{\rm o}(\nu_{\rm o})$, at frequency $\nu_{\rm o}$ with respect to a static observer at spatial infinity. The observed flux of radiation can be calculated by integrating the specific intensity over the observer's local sky. Using Liouville's theorem,  $I_{\rm o} = (\nu_{\rm o}/\nu_{\rm e})^3 I_{\rm e}$, and the observed flux of radiation can be written in terms of the specific intensity with respect to the rest frame of the emitter, $I_{\rm e}$, as
\be\label{eq-thin-Fobs}
F_{\rm o} (\nu_{\rm o}) = \int \frac{g^3}{D^2} I_{\rm e}(\nu_{\rm e}, r_{\rm e}, \vartheta_{\rm e})\, dX dY \, ,
\ee
where $X$ and $Y$ are the Cartesian coordinates of the image of the disk projected on the local sky of the distant observer, $r_{\rm e}$ is the radius of emission in the disk, $\vartheta_{\rm e}$ is the emission angle (i.e., the angle between the normal to the disk and the photon propagation direction in the rest-frame of the gas), $D$ is the distance of the observer from the source, and $g$ is the redshift factor defined by
\be\label{eq:redshift}
g = \frac{\nu_{\rm o}}{\nu_{\rm e}} = \frac{(k_{a}u^{a})_{\rm o}}{(k_{b}u^{b})_{\rm e}}\, .
\ee
Here, $k_{a}$ is the canonical conjugate momentum of an emitted photon and $u^{a}_{\rm o}$ and $u^{a}_{\rm e}$ are the four velocities of the observer and emitter, respectively. Since the system under consideration is stationary and axisymmetric, it admits two Killing vectors corresponding to $\partial_t$ and $\partial_\phi$. Noether's theorem suggests the existence of two conserved quantities for the particles in the disk, $E=-p_t$ and $L_z=p_\phi$.  From the definition of $E$ and $L_z$, we find the expression for two components of the four velocity, $u^t$ and $u^{\rm \phi}$, in terms of the constants of motion
\be
\dot t &=& -\frac{Eg_{\phi\phi}+L_{z}g_{t\phi}}{g_{tt}g_{\phi\phi}-g_{t\phi}^{2}}, \label{eq:dott}
\\
\dot\phi &=& \frac{Eg_{t\phi}+L_{z}g_{tt}}{g_{tt}g_{\phi\phi}-g_{t\phi}^{2}}, \label{eq:dotphi}
\ee
where the dot represents a derivative with respect to the affine parameter. Substitution in the four velocity normalization condition $(u^a u_a=-1)$ yields
\begin{equation}
g_{rr}\dot r^{2}+g_{\theta\theta}\dot\theta^{2}=V_{\text{eff}}(r,\theta;E,L_{z}),
\end{equation}
where the effective potential is defined as 
\begin{equation}
V_{\text{eff}}=-1-\frac{E^{2}g_{\phi\phi}+2EL_{z}g_{t\phi}+L_{z}^{2}g_{tt}}{g_{tt}g_{\phi\phi}-g_{t\phi}^{2}}.\label{eq:Veff}
\end{equation}
Since we are interested in the geodesic motion of the particles in the equatorial plane ($\theta=\pi/2$ and $\dot \theta=0$) with circular orbits ($\dot r=0$ and $\ddot r=0$), we can impose the additional constraints, $V_{\text{eff}}=0$ and $\partial V_{\text{eff}}/\partial r=0$, to solve for $E$ and $L_z$ as
\be
E&=&-\left( g_{tt}+g_{t\phi}\omega \right)\dot{t} \nonumber\\
&=& -\frac{g_{tt}+g_{t\phi}\omega}{\sqrt{-(g_{tt}+2g_{t\phi}\omega+g_{\phi\phi}\omega^{2})}},\label{eq:E} 
\\
L_{z}&=& \left( g_{t\phi}+g_{\phi\phi}\omega \right)\dot{t} \nonumber\\
&=&\frac{g_{t\phi}+g_{\phi\phi}\omega}{\sqrt{-(g_{tt}+2g_{t\phi}\omega+g_{\phi\phi}\omega^{2})}},\label{eq:Lz}
\ee
where the angular velocity of the particles in the disk is given by
\begin{equation}
\omega=\frac{d\phi}{dt}=\frac{-g_{t\phi,r}\pm\sqrt{(g_{t\phi,r})^{2}-g_{tt,r}g_{\phi\phi,r}}}{g_{\phi\phi,r}},\label{eq:angvel}
\end{equation}
and
\begin{equation}
\dot t = \frac{1}{\sqrt{-(g_{tt}+2g_{t\phi}\omega+g_{\phi\phi}\omega^{2})}}.\label{eq:tdot}
\end{equation}

The motion of the photons can similarly be characterised by two constants, $E^{\gamma}$ and $L_z^{\gamma}$, and hence the four momentum for the photons could be written as  $k_{a}=(-E^{\gamma},k_{r},k_{\theta},L_{z}^{\gamma})$.
The geodesic equations for the photons can be written by exploiting the symmetries of the system as 
\begin{align}
\frac{dt}{d\lambda'} =& -\frac{bg_{t\phi} + g_{\phi\phi}}{g_{tt}g_{\phi\phi}-g_{t\phi}^{2}}, \label{eq:dt}
\\
\frac{d\phi}{d\lambda'} =& b\frac{g_{t\phi}+g_{tt}}{g_{tt}g_{\phi\phi}-g_{t\phi}^{2}}, \label{eq:dp}
\end{align}
where $\lambda'\equiv E/\lambda$ is the normalized affine parameter and $b\equiv L_{z}^{\gamma}/E^{\gamma}$ is the impact parameter. The $r$ and $\theta$ components of the geodesic equation are given by
\begin{widetext}
\begin{align}
\frac{d^{2}r}{d\lambda'^{2}}=&-\Gamma^{r}_{tt}\left(\frac{dt}{d\lambda'}\right)^{2}-\Gamma^{r}_{rr}\left(\frac{dr}{d\lambda'}\right)^{2}-\Gamma^{r}_{\theta\theta}\left(\frac{d\theta}{d\lambda'}\right)^{2}-\Gamma^{r}_{\phi\phi}\left(\frac{d\phi}{d\lambda'}\right)^{2}-2\Gamma^{r}{t\phi}\left(\frac{dt}{d\lambda'}\right)\left(\frac{d\phi}{d\lambda'}\right)-2\Gamma^{r}_{r\theta}\left(\frac{dr}{d\lambda'}\right)\left(\frac{d\theta}{d\lambda'}\right), \label{eq:d2r}
\\
\frac{d^{2}\theta}{d\lambda'^{2}}=&-\Gamma^{\theta}_{tt}\left(\frac{dt}{d\lambda'}\right)^{2}-\Gamma^{\theta}_{rr}\left(\frac{dr}{d\lambda'}\right)^{2}-\Gamma^{\theta}_{\theta\theta}\left(\frac{d\theta}{d\lambda'}\right)^{2}-\Gamma^{\theta}_{\phi\phi}\left(\frac{d\phi}{d\lambda'}\right)^{2}-2\Gamma^{\theta}{t\phi}\left(\frac{dt}{d\lambda'}\right)\left(\frac{d\phi}{d\lambda'}\right)-2\Gamma^{\theta}_{r\theta}\left(\frac{dr}{d\lambda'}\right)\left(\frac{d\theta}{d\lambda'}\right), \label{eq:d2th}
\end{align}
\end{widetext}
where $\Gamma^{a}_{bc}$ are the Christoffel symbols of the metric.

The four velocity of the particles in the disk on the other hand can be given from Eq.~\ref{eq:angvel} and Eq.~\ref{eq:tdot} as
\begin{equation}
u_{\rm e}^{a}=\dot t(1,0,0,\omega),
\end{equation}
For the static observer at infinity, the four velocity is given by $u^a_{\rm o}=(1,0,0,0)$. Substituting these expressions in Eq.~\ref{eq:redshift}, we find the redshift factor is given by
\begin{equation}
g=\frac{\sqrt{-(g_{tt}+2g_{t\phi}\omega+g_{\phi\phi}\omega^{2})}}{1-\omega b}.
\end{equation}
Following Cunningham \citep{Cunningham:1975zz}, we perform the integration of the specific intensity by introducing a coordinate transformation from the celestial coordinates of the observer $(X, Y)$ to the new set of coordinates $(r_{\rm e}, g^*)$, where $g^*$ is defined as 
\be
g^* = \frac{g - g_{\rm min}}{g_{\rm max} - g_{\rm min}} \, ,
\ee
with $g_{\rm min}(r_{\rm e}, i)$ and $g_{\rm max}(r_{\rm e}, i)$ being the minimum and maximum redshift factor of a photon emitted from radius $r_{\rm e}$ and observed by an observer with inclination angle $i$. This transformation can be easily implemented by introducing the transfer function $f$ given by
\be\label{eq-trf}
f(g^*,r_{\rm e},i) = \frac{1}{\pi r_{\rm e}} g 
\sqrt{g^* (1 - g^*)} \left| \frac{\partial \left(X,Y\right)}{\partial \left(g^*,r_{\rm e}\right)} \right| \, ,
\ee
where, $\left| \frac{\partial \left(X,Y\right)}{\partial \left(g^*,r_{\rm e}\right)} \right|$, the Jacobian, is calculated using a general relativistic ray-tracing code. The expression for the observed flux in Eq.~\ref{eq-thin-Fobs} can now be written as
\be\label{eq-Fobs}
F_{\rm o} (\nu_{\rm o}) 
&=& \frac{1}{D^2} \int_{r_{\rm in}}^{r_{\rm out}} \int_0^1
\pi r_{\rm e} \frac{ g^2}{\sqrt{g^* (1 - g^*)}} f(g^*,r_{\rm e},i) \nonumber \\
&& I_{\rm e}(\nu_{\rm e},r_{\rm e},\vartheta_{\rm e}) \, dg^* \, dr_{\rm e} \, ,
\ee
where $r_{\rm in}$ and $r_{\rm out}$ are the inner and the outer edge of the accretion disk, respectively. If the disk is not truncated, $r_{\rm in}$ corresponds to the radius of the innermost stable circular orbit (ISCO). The ISCO can be found by substituting Eq.~\ref{eq:E} and Eq.~\ref{eq:Lz} in Eq.~\ref{eq:Veff} and solving for the condition $\partial^2V_{\mathrm{eff}}/\partial r^2=0$.


\section{Reflection spectrum for non-geodesic motion}\label{s-ngm}

\subsection{Theoretical framework}

According to the WEP, all freely falling test-particles must follow the geodesics of the same spacetime metric. A violation of the WEP would thus lead to different geodesics for different classes of particles. We implement this behavior in a simple way. We study the violation of the WEP with two simple cases: in the first case, we consider that photons emitted from the disk follow the geodesics of the Kerr background whereas massive particles in the accretion disk follow the geodesics of a deformed Kerr spacetime. In the second case, we consider Kerr geodesics for massive particles in the disk and non-Kerr geodesics for photons. The non-Kerr geodesics are modeled by considering the Johanssen metric, which modifies the Kerr spacetime while retaining a Carter-constant-like symmetry~\cite{Johannsen:2015pca}. The line element of this spacetime, with only the leading-order non-Kerr deformations, in Boyer-Lindquist coordinates is given by 
\be\label{eq:johmet}
ds^2 &=& - \frac{\tilde{\Sigma} \left(\Delta - a^2 A_2^2 \sin^2\theta\right)}{B^2} \, dt^2 \nonumber \\
&& - \frac{2 a \left[ \left(r^2 + a^2\right) A_1 A_2 - \Delta\right] 
\tilde{\Sigma} \sin^2\theta}{B^2} \, dt \, d\phi \nonumber\\ 
&& + \frac{\tilde{\Sigma}}{\Delta A_5} \, dr^2
+ \tilde{\Sigma} \, d\theta^2 \nonumber \\
&& + \frac{\left[ \left(r^2 + a^2\right)^2 A_1^2 - a^2 \Delta \sin^2\theta\right] 
\tilde{\Sigma} \sin^2\theta}{B^2} \, d\phi^2
\ee
where $a = J/M$ (the dimensionless spin parameter used later is $a_* = a/M$),
\be
&&B = \left(r^2 + a^2\right) A_1 - a^2 A_2 \sin^2\theta \, , \quad
\tilde{\Sigma} = \Sigma + f \, , \nonumber\\
&&\Sigma = r^2 + a^2 \cos^2\theta \, , \quad
\Delta = r^2 - 2 M r + a^2 \, ,
\ee
and
\be\label{eq-def}
&& f = \epsilon_3 \frac{M^3}{r} \, , \qquad
A_1 = 1 + \alpha_{13} \left(\frac{M}{r}\right)^3 \, , \nonumber\\
&& A_2 = 1 + \alpha_{22} \left(\frac{M}{r}\right)^2 \, , \qquad
A_5 = 1 + \alpha_{52} \left(\frac{M}{r}\right)^2 \, . \qquad
\ee
This metric reduces to the Kerr solution for $\epsilon_3 = \alpha_{13} = \alpha_{22} = \alpha_{52} = 0$, has the correct Newtonian limit and is consistent with the current PPN constraints~\cite{Johannsen:2015pca}.

For this work, as a proof-of-principle to illustrate the capability of X-ray reflection spectroscopy to test the WEP, we will only consider a non-zero value for $\alpha_{13}$ and set all other Johannsen parameters to zero. As has been discussed in past work (see~e.g.~\cite{Riaz:2020svt}) large values of $\alpha_{13}$ lead to pathologies in the spacetime such as a naked singularity or coordinate singularities in Boyer-Lindquist coordinates. However, as will be seen in Sec.~\ref{s-obs}, the constraints we find on $\alpha_{13}$ and departures from the WEP are very strong and do not approach these pathological regions.

\subsection{Non-geodesic motion of massive particles}

\begin{figure}[h!]
\begin{center}
\includegraphics[width=8.0cm,trim={0cm 0cm 0cm 0cm},clip]{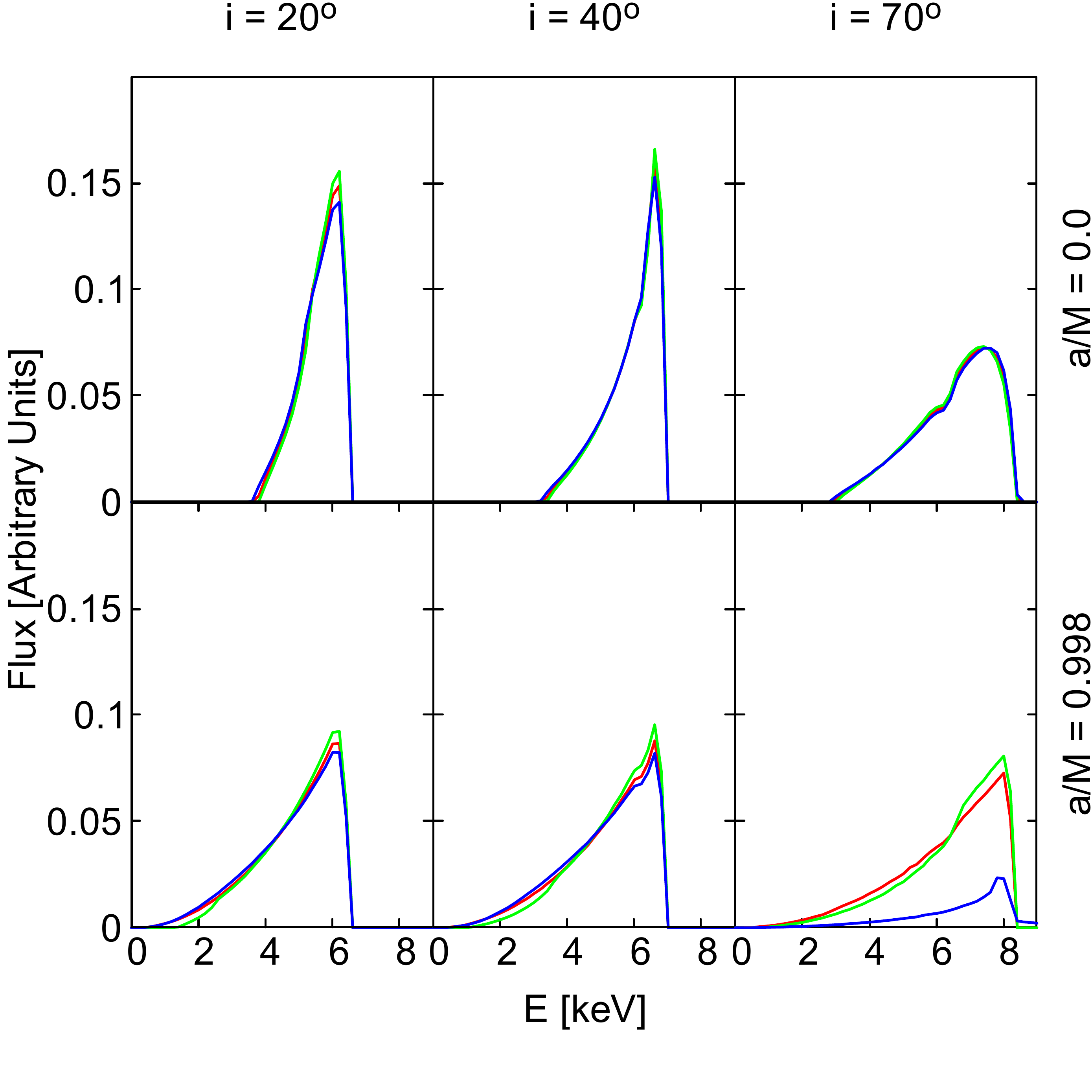}
\end{center}
\vspace{-0.9cm}
\caption{Impact of non-geodesic motion of massive particles on an iron line for different values of the black hole spin parameter and of the inclination angle of the disk with respect to the line of sight of the observer. The parameter of non-geodesic motion is $\beta_{13} = - 1$ (red), 0 (green), and 1 (blue) when the black hole spin parameter is $a_* = 0$ (top) and $\beta_{13} = - 0.35$ (red), 0 (green), and 0.35 (blue) when $a_* = 0.998$ (bottom). In all plots the emissivity profile of the disk is described by a power-law with emissivity index $q = 3$, the iron line is at 6.4~keV in the rest-frame of the gas, and the inner edge of the disk is at the ISCO radius. \label{f-linesb}}
\end{figure}

We first consider non-Kerr geodesics for the massive particles in the accretion disk, while photons follow the standard Kerr geodesics. We introduce the notation $g^{\rm k}_{\mu\nu}$ and $g^{\rm nk}_{\mu\nu}$ to denote the background Kerr metric and the metric given in Eq.~\ref{eq:johmet} used to model the non-Kerr geodesics, respectively. With the introduction of the two metrics, we see that Eqs.~\ref{eq:dott}-\ref{eq:tdot} are modified to contain $g^{\rm nk}_{\mu\nu}$. This, in turn, modifies the ISCO radius, given by solving $\partial^2V_{\mathrm{eff}}/\partial r^2=0$ with $V_{\rm eff}$ defined as 
\begin{equation}
V_{\text{eff}}=-1-\frac{E^{2}g^{\rm nk}_{\phi\phi}+2EL_{z}g^{\rm nk}_{t\phi}+L_{z}^{2}g^{\rm nk}_{tt}}{g^{\rm nk}_{tt}g^{\rm nk}_{\phi\phi}-(g^{\rm nk}_{t\phi})^{2}}.\label{eq:Veff2}
\end{equation}
Similarly, the redshift factor is modified to reflect the non-Kerr geodesics of the particles in the disk and the expression is given by 
\begin{equation}
g=\frac{\sqrt{-(g^{\rm nk}_{tt}+2g^{\rm nk}_{t\phi}\omega^{\rm nk}+g^{\rm nk}_{\phi\phi}(\omega^{\rm nk})^{2})}}{1-\omega^{\rm nk} b},
\end{equation}
where $\omega^{\rm nk}$ denotes the angular velocity of the particles with the non-Kerr metric in Eq.~\ref{eq:angvel}. Since the photons still follow the geodesics of the Kerr background, the differential equations determining the geodesics of the photons, Eqs.~\ref{eq:dt}-\ref{eq:d2th}, used in the ray-tracing code to calculate the transfer functions remains the same. We plot the impact of non-geodesic motion of massive particles on an iron line in Fig.~\ref{f-linesb}. For convenience, we introduce the parameters $\beta_{13}$ and $\gamma_{13}$ in place of the deformation parameter $\alpha_{13}$ for the massive particles and photons, respectively.

The general trend in the iron line shows an increasing deviation from the Kerr case both with an increase in the spin parameter and the inclination angle. For higher values of the spin parameter, the inner edge of the accretion disk, which is set at the ISCO radius, is closer to the black hole and thus the iron line profile is more sensitive to variations of the gravitational field. For higher inclination angles, Doppler boosting is increased so we are more sensitive to the orbital motion of the gas in the disk.

\subsection{Non-geodesic motion of photons}

\begin{figure}[h!]
\begin{center}
\includegraphics[width=8.0cm,trim={0cm 0cm 0cm 0cm},clip]{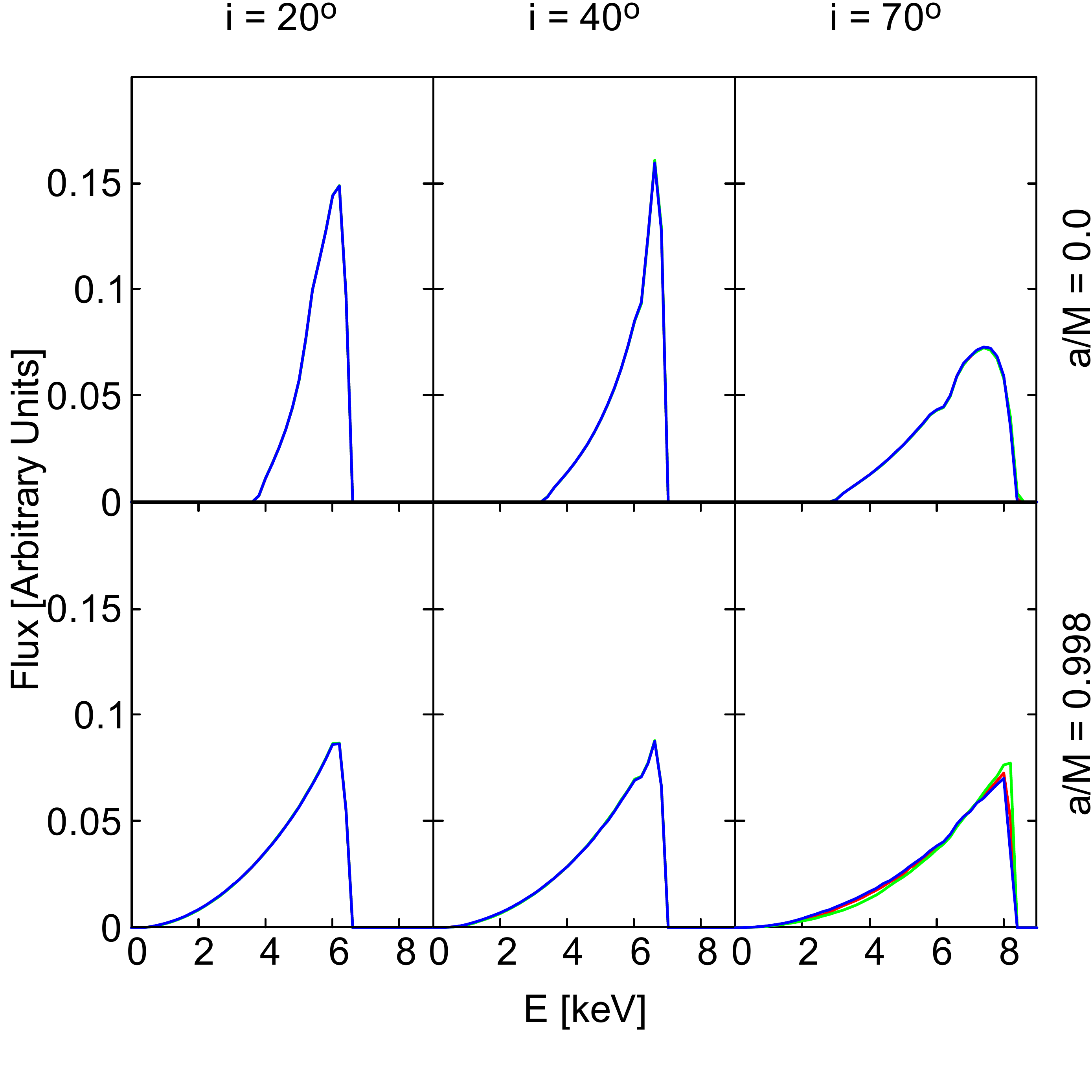}
\end{center}
\vspace{-0.9cm}
\caption{As in Fig.~\ref{f-linesb} for the case of non-geodesic motion of photons. The parameter of non-geodesic motion is $\gamma_{13} = - 1$ (red), 0 (green), and 1 (blue) when the black hole spin parameter is $a_* = 0$ (top) and $\gamma_{13} = - 0.35$ (red), 0 (green), and 0.35 (blue) when $a_* = 0.998$ (bottom).\label{f-linesc}}
\end{figure}

In the second case, we consider that photons follow the geodesics of a non-Kerr metric whereas the massive particles follow the geodesics of the Kerr background. The differential equations used in the ray tracing code, Eqs.~\ref{eq:dt}-\ref{eq:d2th}, for the geodesics of the photons are modified to use the non-Kerr Johanssen metric $g^{\rm nk}_{\mu\nu}$. The equations determining the motion of the massive particles on the other hand, Eqs.~\ref{eq:dott}-\ref{eq:tdot}, stay the same. We plot the impact of non-geodesic motion of photons on an iron line in Fig.~\ref{f-linesc}.

As we could expect, the impact of possible non-geodesic motion of photons on an iron line profile is weaker than the previous case of non-geodesic motion of the massive particles in the disk. Here the ISCO radius is that of the Kerr metric, as well as the structure of the accretion disk. Only light bending is modified and this explains why the effect of a possible non-geodesic motion of photons is more important for large viewing angles.

\section{Observational constraints}\label{s-obs}

The framework described in the previous section with two metrics, one for massive particles and one for photons, is implemented in the reflection model {\tt relxill\_nk}~\cite{Bambi:2016sac,Abdikamalov:2019yrr,Abdikamalov:2020oci}. All the details of the new framework to test the WEP are encoded in the FITS file of the relativistic convolution model, while all other parts of the model are the same as the standard {\tt relxill\_nk}. With this new model, we can observationally constrain $\beta_{13}$ and $\gamma_{13}$ by analyzing relativistic reflection features in the spectra of accreting black holes.

Here we choose a 2019 \textsl{NuSTAR} observation of the black hole binary EXO~1846--031. Previous studies of these data in Kerr and non-Kerr spacetimes have shown that the stellar-mass black hole in EXO~1846--031 has a spin parameter $a_*$ close to 1 and that the inclination angle of the disk is very high, which are the two key-ingredients found in the previous section to maximize the impact of non-geodesic motion on the shape of an iron line. The source is known to have very simple spectra with no intrinsic absorptions and display strong relativistic reflection features. \textsl{NuSTAR} is currently the most suitable observatory to study reflection features of bright sources because it has no pile-up problem to observe bright X-ray binaries and because of the wide energy band of its instruments, from 3 to 79~keV.

\subsection{Source}

EXO~1846--031 is a low mass X-ray binary, first observed by \textsl{EXOSAT} on 1985~April~3~\cite{1985IAUC.4051....1P} and confirmed in 1993 \cite{1993A&A...279..179P}. Since its discovery, three outbursts have been observed: the first one in 1985 (when it was discovered), the second one in 1994 observed by \textsl{CGRO}/BATSE~\cite{1994IAUC.6096....1Z}, and the third and last outburst was in July 2019 and first detected by \textsl{MAXI}~\cite{2019ATel12968....1N}. The 2019 outburst was then observed by multiple instruments. In this work, we analyze a \textsl{NuSTAR} spectrum of the 2019 outburst. This observation was first analyzed in \cite{Draghis:2020ukh} for measuring the black hole spin (assuming the Kerr metric).

\subsection{Data Reduction}

\textsl{NuSTAR}~\cite{Harrison:2013md} observed EXO~1846--031 on 2019~August~29 for about 22~ks with both its detectors, Focal Plane Module (FPM) A and B. The instrumental unprocessed data were downloaded from the HEASARC website archives and converted to processed ones using NUSTARDAS (distributed as part of the HEASOFT package) and CALDB version 20200912. The processed event files were used to obtain source and background region of 180~arcsec taken around the source and far away from it on the same detector, respectively. The spectra, response, and ancillary files were generated using the NUPRODUCTS routine of NUSTARDAS. Finally, the source spectra are binned to 30 counts per bin in order to use the $\chi^2$ statistic.

\begin{figure*}[t]
\begin{center}
\includegraphics[width=8.5cm,trim={2.5cm 0cm 2.5cm 18cm},clip]{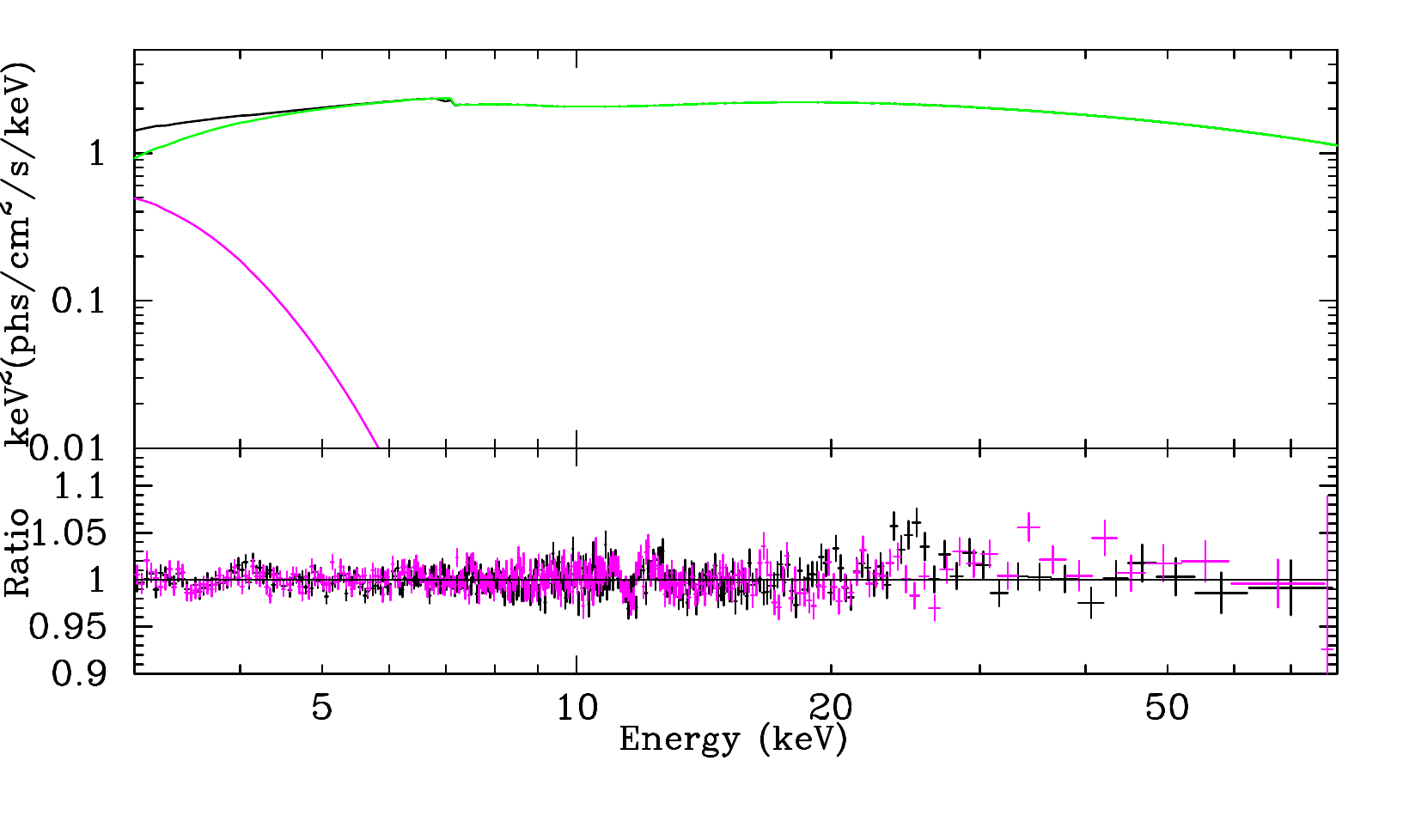} \hspace{0.2cm}
\includegraphics[width=8.5cm,trim={2.5cm 0cm 2.5cm 18cm},clip]{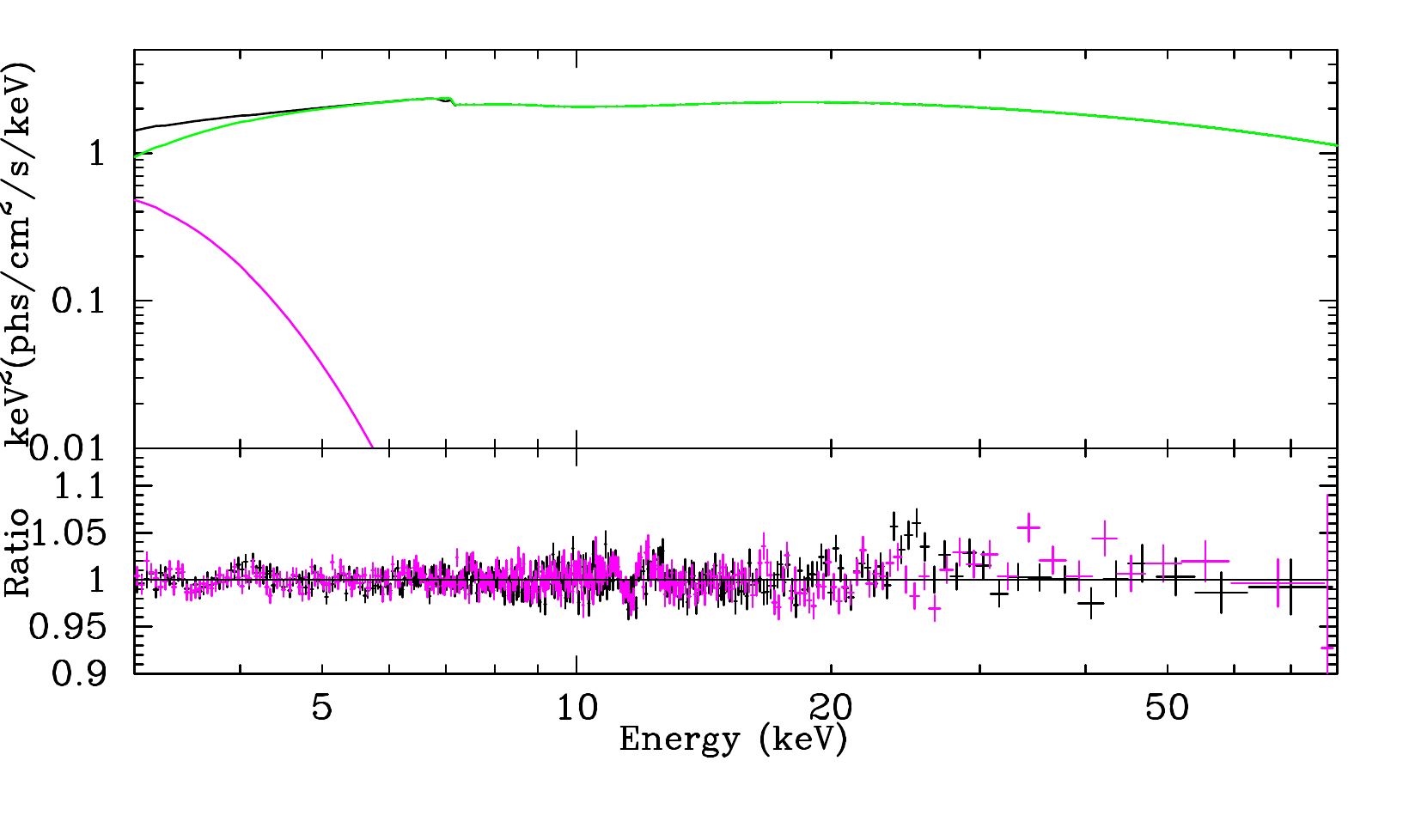}
\end{center}
\vspace{-0.8cm}
\caption{The best-fit model (upper quadrants) and data to best-fit model ratio (lower quadrants) for the case of non-geodesic motion of massive particles (left panel) and of photons (right panel). In the upper quadrants, the black curve refers to the total model, the green curve to the {\tt relxill\_nk} component (relativistic reflection spectrum), and the magenta curve to the {\tt diskbb} component (thermal spectrum). In the lower quadrants, black data are for FPMA and magenta data for FPMB.  
\label{f-ratio}}
\end{figure*}

\subsection{Spectral Analysis}  

In our analysis, we use both the FPMA and FPMB data throughout the energy range of 3-79~keV and we follow Ref.~\cite{Tripathi:2020yts}. If we fit the data with an absorbed power-law, we see a very broad iron line in the residuals. We thus fit the data including a relativistic reflection component. In XSPEC language, the model is {\tt tbabs}$\times${\tt relxill\_nk}, where {\tt tbabs} describes the Galactic absorption~\cite{Wilms:2000ez} and {\tt relxill\_nk} describes the power-law continuum from the corona and the reflection spectrum from the disk. We analyze the data with three models:
\begin{enumerate}
\item Model~0: $\beta_{13} = \gamma_{13} = 0$ (general relativity, all particles follow the geodesics of the Kerr metric).
\item Model~1: $\beta_{13}$ free and $\gamma_{13} = 0$ (massive particles may not follow the geodesics of the Kerr metric, photons do).
\item Model~2: $\beta_{13} = 0$ and $\gamma_{13}$ free (massive particles follow the geodesics of the Kerr metric, photons may not).
\end{enumerate}

In {\tt tbabs}, the free parameter is the hydrogen column density $N_{\rm H}$. In {\tt relxill\_nk}, the continuum from the corona is specified by the photon index $\Gamma$ and the high-energy cutoff $E_{\rm cut}$. In the reflection spectrum from the disk, the emissivity profile is described by a broken power-law and we have thus three free parameters: the inner emissivity index $q_{\rm in}$, the outer emissivity index $q_{\rm out}$, and the breaking radius $R_{\rm br}$. The accretion disk is also characterized by the ionization parameter $\xi$ (measured in units of erg~cm~s$^{-1}$), the iron abundance $A_{\rm Fe}$ (in units of the solar iron abundance), and its viewing angle $i$, namely the angle between the normal to the disk and our line of sight. The spacetime metric is specified by the black hole spin parameter $a_*$ and the deformation parameter, $\beta_{13}$ in Model~1 and $\gamma_{13}$ in Model~2\footnote{We note that the current version of the model does not permit us to consider simultaneously free $\beta_{13}$ and $\gamma_{13}$.}. The inner edge of the disk is set at the ISCO radius, so it depends on $a_*$ and the deformation parameters.

After fitting the data with the above model, there are residuals at low energies and an absorption feature around 7~keV. We repeat the fits including {\tt diskbb} (thermal spectrum of the accretion disk) to fit the residuals at low energies and a Gaussian line for the absorption feature at 7~keV. Fig.~\ref{f-ratio} shows the best fit model and the data to best fit model ratio for Model~1 (left panel) and Model~2 (right panel).

\begin{figure*}[t]
\begin{center}
\includegraphics[width=17cm,trim={0cm 0cm 0cm 0cm},clip]{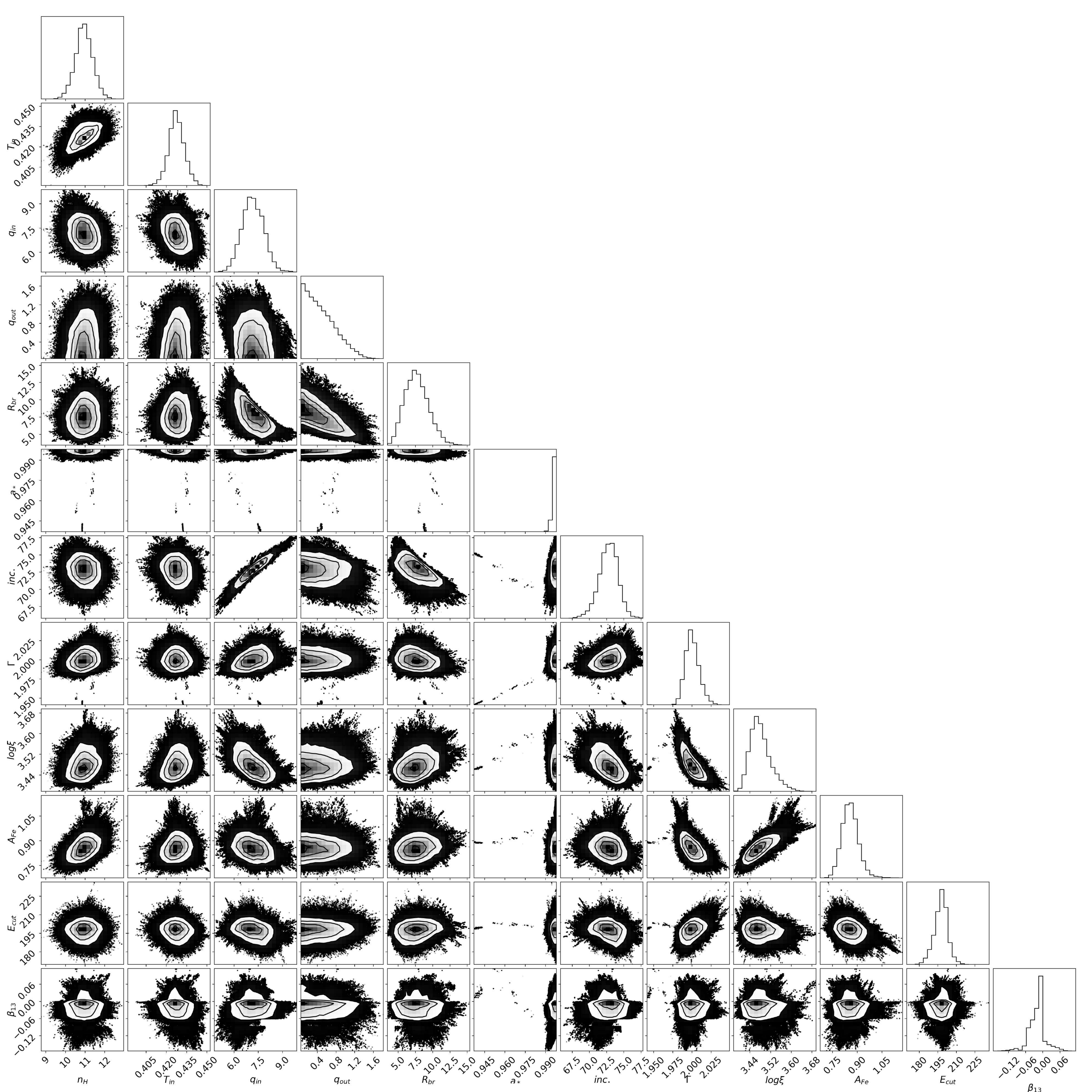}
\end{center}
\vspace{-0.5cm}
\caption{Corner plot for the main free parameter-pairs in the scenario of possible deviation from geodesic motion by massive particles (Model~1) after the MCMC run. The 2D plots show the 1-, 2- and 3-$\sigma$ confidence contours. \label{f-mcmcb}}
\end{figure*}

\begin{figure*}[t]
\begin{center}
\includegraphics[width=17cm,trim={0cm 0cm 0cm 0cm},clip]{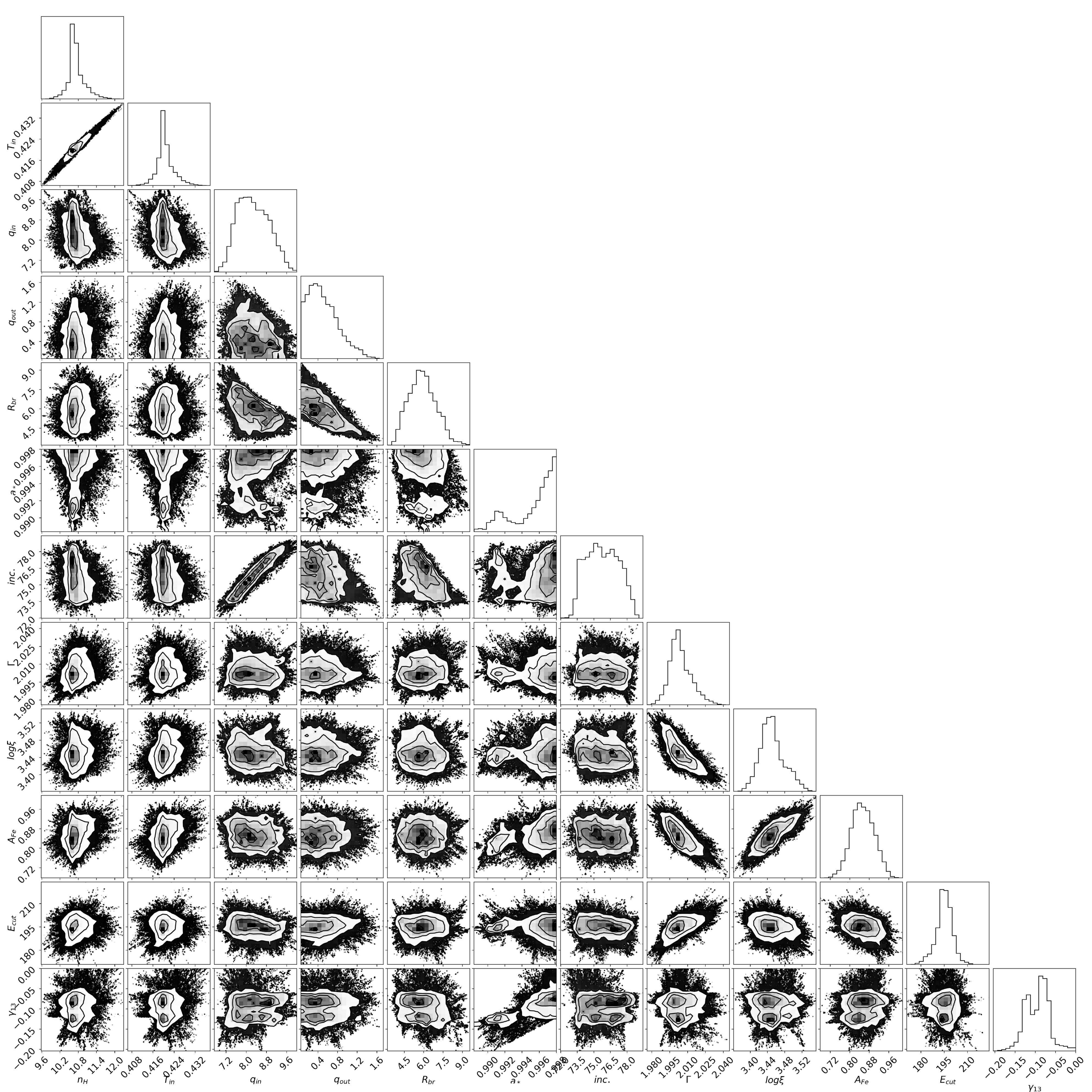}
\end{center}
\vspace{-0.5cm}
\caption{As in Fig.~\ref{f-mcmcb} in the scenario of possible deviation from geodesic motion by photons (Model~2). \label{f-mcmcc}}
\end{figure*}

We use the python script of Jeremy Sanders\footnote{The script is available at 
\href{https://github.com/jeremysanders/xspec_emcee}{https://github.com/jeremysanders/xspec\_emcee}.} to run Markov Chain Monte Carlo (MCMC) simulations of the best fit model for both $\beta_{13}$ and $\gamma_{13}$ in order to estimate better the uncertainties of the parameters and explore their correlations. We run MCMC for 20000 iterations and 300 walkers, burning the initial 1000 iterations. The total number of samples are $6\times 10^6$ steps. The statistical error is calculated for the 90\% confidence interval across the whole parameter chain and is shown in Tab.~\ref{tab1}. The $\chi^2$ reported in the table are calculated using $\chi^2$ statistics. The corner plots for Model~1 and Model~2 are reported in Fig.~\ref{f-mcmcb} and Fig.~\ref{f-mcmcc}, respectively.

\begin{table}
\centering
\renewcommand\arraystretch{1.35}
\begin{tabular}{lccc}
\hline\hline
Parameter & Model~0 & Model~1 & Model~2 \\
\hline
{\tt tbabs} \\
$N_{\rm H}$ [$10^{22}$~cm$^{-2}$] & $10.7^{+0.6}_{-0.7}$ & $10.9^{+0.7}_{-0.7}$ & $10.7^{+0.6}_{-0.4}$ \\
\hline
{\tt relxill\_nk} \\
$q_{\rm in}$ & $8.3^{+0.2}_{-1.5}$ & $7.1^{+1.1}_{-1.0}$ & $8.2^{+1.1}_{-0.9}$ \\
$q_{\rm out}$ & $<1.2$ & $0.99^{+0.08}_{-0.92}$ & $0.5^{+0.7}_{-0.4}$ \\
$R_{\rm br}$ [$r_{\rm g}$] & $6.8^{+0.7}_{-2.8}$ & $7.6^{+3.1}_{-2.5}$&   $5.8^{+1.8}_{-1.6}$ \\
$a_*$ & $0.998^{}_{-0.003}$ & $0.997^{\rm +(P)}_{-0.003}$ & $0.996^{\rm + (P)}_{-0.006}$ \\
$i$ [deg] & $75.0^{+0.6}_{-0.4}$ & $72.8^{+2.1}_{-1.5}$ & $76.3^{+2.4}_{-2.1}$ \\
$\Gamma$ & $2.000^{+0.021}_{-0.049}$ & $1.999^{+0.017}_{-0.012}$ & $2.001^{+0.019}_{-0.012}$ \\
$\log\xi$ [erg~cm~s$^{-1}$] & $3.45^{+0.03}_{-0.04}$ & $3.48^{+0.09}_{-0.04}$ & $3.45^{+0.05}_{-0.04}$ \\
$A_{\rm Fe}$ & $0.86^{+0.05}_{-0.09}$ & $0.86^{+0.08}_{-0.08}$ & $0.85^{+0.08}_{-0.07}$ \\
$\beta_{13}$ & $0^\star$ & $-0.02^{+0.04}_{-0.04}$& $0^\star$ \\
$\gamma_{13}$ & $0^\star$ & $0^\star$ & $-0.09^{+0.05}_{-0.04}$ \\
$E_{\rm cut}$ [keV] & $194^{+7}_{-9}$ & $198^{+8}_{-10}$ & $195^{+8}_{-10}$ \\
\hline
{\tt diskbb} \\
$T_{\rm in}$ [keV] & $0.423^{+0.007}_{-0.013}$ & $0.428^{+0.007}_{-0.013}$ & $0.420^{+0.007}_{-0.005}$ \\
\hline
{\tt gauss} \\
$E_{\rm line}$ [keV] & $7.00^{+0.05}_{-0.05}$ & $7.00^{+0.05}_{-0.05}$ & $7.01^{+0.06}_{-0.06}$ \\
\hline
$\chi^2$/dof & 2751.89/2595 & 2751.84/2594 & 2749.55/2594 \\
& =1.06046 & =1.06085 & =1.05997 \\
\hline\hline
\end{tabular}
\caption{Summary of the best-fit values of Model~0 (general relativity), Model~1 (possible non-geodesic motion of the particles of the accretion disk), and Model~2 (possible non-geodesic motion of photons). The reported uncertainties correspond to the 90\% confidence level for one relevant parameter ($\Delta\chi^2 = 2.71$). $^\star$ indicates that the value of the parameter is frozen in the fit. (P) means that the 90\% confidence level reaches the upper boundary of the black hole spin parameter ($a_*^{\rm max} = 0.998$). See the text for more details. }
\label{tab1}
\end{table}

\section{Discussion and conclusions}\label{s-dis}

In this paper, we have proposed to test the WEP in the strong gravity region around black holes by analyzing the reflection features of their accretion disks. We have analyzed a \textsl{NuSTAR} observation of the black hole binary EXO~1846--031 with three models: general relativity in which the WEP holds (Model~0), a model in which massive particles may have non-geodesic motion (Model~1), and a model in which photons may show non-geodesic motion (Model~2).

Our results are consistent with the WEP. As we can see from Tab.~\ref{tab1}, the estimate of $\beta_{13}$ in Model~1 is consistent with 0 at the 90\% confidence level. For Model~2, the estimate of $\gamma_{13}$ reported in the table is not consistent with 0 at the 90\% confidence level, but this is because there are several minima and the measurement reported in Tab.~\ref{tab1} refers to the minimum with the lowest value of $\chi^2$. If we compare the $\chi^2$ of Model~0 and Model~2, we see that $\Delta\chi^2 = 2.34$, which is less than $\Delta\chi^2 = 2.71$ corresponding to the 90\% confidence limit.

Parameter degeneracy is quite a common issue when we fit the data of accreting black holes, even when we assume the Kerr metric. This is because the models used to fit the data normally have many free parameters. We would like to point out that this is not a particular problem for our analysis: as we can see from the corner plots in Fig.~\ref{f-mcmcb} and Fig.~\ref{f-mcmcc}, the deformation parameters $\beta_{13}$ and $\gamma_{13}$ do not show any strong correlation with the other model parameters. We only see a modest correlation between $\gamma_{13}$ and the black hole spin parameter $a_*$. However, such a favorable situation is due to the choice of the source. The black hole in EXO~1846--031 is rotating very fast, so the inner part of the disk enters the strong gravity region and this maximizes the relativistic effects in the reflection spectrum of the source. If we had chosen a slowly-rotating source, we would have certainly found a more pronounced correlation between the deformation parameters and some of the other parameters of the model. In particular, in the case of slowly-rotating black holes there is a strong degeneracy between $\beta_{13}$ and the black hole spin $a_*$, since both parameters determine the value of the ISCO radius and other relativistic effects are too weak in a slowly-rotating object.

In Tab.~\ref{tab1}, the uncertainties on the model parameters refer to the statistical uncertainties only, ignoring the systematic uncertainties related to our theoretical model. However, for this specific observation of \textsl{NuSTAR} the systematic uncertainties are well under control (see list below) and we think that our measurement is quite robust.

{\it Inner edge of the disk}. In our analysis, we assumed that the inner edge of the disk is at the ISCO radius. While it is still a controversial issue if this is the case when a source is in the hard state (which is the spectral state of EXO~1846--031 in our \textsl{NuSTAR} observation), see \cite{Bambi:2020jpe} and references therein, such an assumption does not have any implication on our results because the fit prefers a spin parameter at, or very close to, the maximum value allowed by the model. Relaxing this assumption, we would obtain the same results.

{\it Thickness of the disk}. The model used to fit the \textsl{NuSTAR} observation employs an infinitesimally thin disk, while the disk should have a finite thickness, which can be expected to increase as the black hole mass accretion rate increases. However, since the black hole spin is very high, the radiative efficiency is very high too, and the thickness of the disk is modest. These \textsl{NuSTAR} data were fit with a model with a disk of finite thickness in Ref.~\cite{Tripathi:2021wap} without improving the fit and without finding different estimates of the model parameters.

{\it Radiation from the plunging region}. The plunging region, namely the region between the inner edge of the disk and the event horizon of the black hole, is thought to be optically thick for reasonable values of the mass accretion rate~\cite{Reynolds:1997ek}. The radiation from the corona should thus illuminate the plunging region and generate a reflection spectrum, which is instead ignored in our calculations. However, the gas is highly ionized in the plunging region~\cite{Reynolds:1997ek,Wilkins:2020pgu} and, for fast-rotating black holes, the plunging region is very small. As shown in Ref.~\cite{Cardenas-Avendano:2020xtw}, in such a case the radiation from the plunging region has a negligible impact on the estimate of the black hole spin or of the deformation parameters.

{\it Constant ionization parameter}. The ionization parameter $\xi$ is defined as $\xi = F_X/(4\pi n_{\rm e})$, where $F_X$ is the X-ray flux from the corona and $n_{\rm e}$ is the disk electron density. In general, we should expect that $F_X$, $n_{\rm e}$, and $\xi$ are functions of the radial coordinate $r$, while the model used to analyze the observation of EXO~1846--031 assumes a constant ionization parameter over the whole disk. The \textsl{NuSTAR} observation of EXO~1846--031 was analyzed with a model with a non-vanishing ionization gradient in Ref.~\cite{Abdikamalov:2021rty}. No significant difference in the estimate of most model parameters was found.

{\it Electron density in the disk}. In the analysis presented in the previous section, we assumed the default value for the electron density in the disk, $n_{\rm e} = 10^{15}$~cm$^{-3}$. Such a value is likely too low for the accretion disk of stellar-mass black holes~\cite{Jiang:2019xqn,Jiang:2019ztr}. The impact of the electron density on the estimate of the deformation parameters was investigated in Refs.~\cite{Tripathi:2020dni,Zhang:2019ldz}, concluding that the value of $n_{\rm e}$ does not affect the measurement of the deformation parameter, which is consistent with the conclusions in Refs.~\cite{Jiang:2019xqn,Jiang:2019ztr} that the value of $n_{\rm e}$ does not have a significant impact on the determination of the black hole spin parameter when we assume the Kerr metric.

{\it Returning radiation}. A fraction of the radiation emitted from the inner part of the accretion disk returns to the disk because of the strong light bending near the black hole and this generates a secondary reflection spectrum. In general, if the fitting model ignores the returning radiation, the estimate of some model parameters may be biased. However, this may be a problem if the source is observed from a low viewing angle~\cite{Riaz:2020zqb}. For high viewing angles, which is the case of EXO~1846--031, such an effect should be weak and, presumably, negligible~\cite{Riaz:2020zqb}.

In conclusion, X-ray reflection spectroscopy can test the WEP in the strong gravity region around black holes and verify whether massive and/or massless particles follow the geodesics of the Kerr spacetime. Up to now, the WEP has been extensively tested in Earth's gravitational field, without finding any deviation. Tests of the WEP near black holes explore a completely different regime that is worth probing. We note that gravitational waves cannot test whether photons follow the geodesics of the spacetime, and this shows how electromagnetic and gravitational wave tests can be complementary to test general relativity in the next decades.


\vspace{0.5cm}

{\bf Acknowledgments --}
This work was supported by the Innovation Program of the Shanghai Municipal Education Commission, Grant No.~2019-01-07-00-07-E00035, the National Natural Science Foundation of China (NSFC), Grant No.~11973019, and Fudan University, Grant No.~JIH1512604.
R.R. also acknowledges the support from the Shanghai Government Scholarship (SGS).
D.A. is supported through a Teach@T{\"u}bingen Fellowship.


\end{document}